\documentclass[amsmath,nobibnotes,aps,superscriptadress,amssymb,pra,aps,showpacs,superscriptaddress,twocolumn]{revtex4-1}
\usepackage{amsmath,amsfonts,amssymb,amsthm,graphics,graphicx,epsfig,bbm}
\usepackage[colorlinks=true,citecolor=blue,linkcolor=blue,urlcolor=blue]{hyperref}
\usepackage[usenames]{color}
\usepackage{graphicx}
\usepackage{subfigure}
\usepackage{amsmath}
\usepackage{epsfig}
\usepackage{dcolumn}
\usepackage{bm}
\usepackage{color}
\usepackage{times}
\usepackage{epstopdf}
\usepackage{amssymb}
\usepackage{amstext}
\usepackage{latexsym}
\usepackage{hyperref}
\usepackage{amsfonts}
\usepackage{psfrag}
\usepackage{soul,xcolor}
\usepackage[normalem]{ulem}
\usepackage{dsfont}
\usepackage{txfonts}
\usepackage{physics}
\usepackage{footnote}
\usepackage{blindtext}
\usepackage{bbm}
\usepackage{tikz}
\usetikzlibrary{quantikz}
\usepackage{bbold}

\begin{document}
\setstcolor{red}

\title{Digital-analog co-design of the Harrow-Hassidim-Lloyd algorithm}

\author{Ana Martin}
\email{Corresponding author: ana.martinf@ehu.eus}
\affiliation{Department of Physical Chemistry, University of the Basque Country UPV/EHU, Apartado 644, 48080 Bilbao, Spain}
\affiliation{EHU Quantum Center, University of the Basque Country UPV/EHU, Bilbao, Spain}
\affiliation{Quantum Mads, Uribitarte Kalea 6, 48001 Bilbao, Spain}

\author{Ruben Ibarrondo}
\affiliation{Department of Physical Chemistry, University of the Basque Country UPV/EHU, Apartado 644, 48080 Bilbao, Spain}
\affiliation{EHU Quantum Center, University of the Basque Country UPV/EHU, Bilbao, Spain}

\author{Mikel Sanz}
\email{Corresponding author: mikel.sanz@ehu.es}
\affiliation{Department of Physical Chemistry, University of the Basque Country UPV/EHU, Apartado 644, 48080 Bilbao, Spain}
\affiliation{EHU Quantum Center, University of the Basque Country UPV/EHU, Bilbao, Spain}
\affiliation{Ikerbasque Foundation for Science, Plaza Euskadi 5, 48009 Bilbao, Spain}
\affiliation{BCAM-Basque Center for Applied Mathematics, Mazarredo 14, 48009 Bilbao, Basque Country, Spain}

\begin{abstract}
The Harrow-Hassidim-Lloyd quantum algorithm was proposed to solve linear systems of equations $A\vec{x} = \vec{b}$ and it is the core of various applications. However, there is not an explicit quantum circuit for the subroutine which maps the inverse of the problem matrix $A$ into an ancillary qubit. This makes challenging the implementation in current quantum devices, forcing us to use hybrid approaches. Here, we propose a systematic manner to implement this subroutine, which can be adapted to other functions $f(A)$ of the matrix $A$, we present a co-designed quantum processor which reduces the depth of the algorithm, and we introduce its digital-analog implementation. The depth of our proposal scales with the precision $\epsilon$ as $\mathcal{O}(\epsilon^{-1})$, which is bounded by the number of samples allowed for a certain experiment. The co-design of the Harrow-Hassidim-Lloyd algorithm leads to a ``kite-like" architecture, which allows us to reduce the number of required SWAP gates. Finally, merging a co-design quantum processor architecture with a digital-analog implementation contributes to the reduction of noise sources during the experimental realization of the algorithm.
\end{abstract}

\maketitle

\section{Introduction}

\begin{figure*}
\centering
\includegraphics[width=0.98\textwidth]{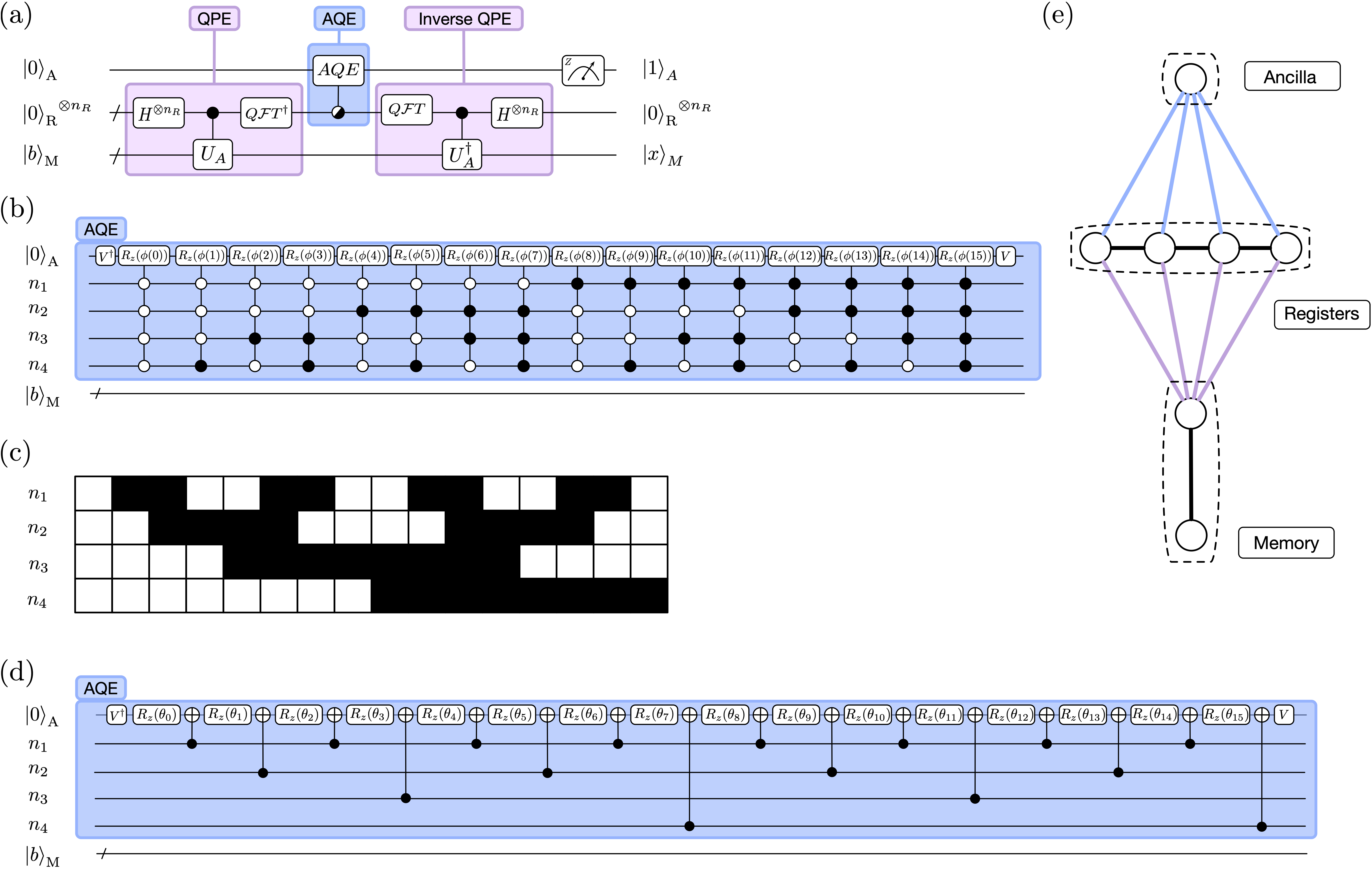}
\caption{{\bf Quantum circuit for the digital implementation of the HHL.} 
{\bf (a) Schematic representation of the HHL algorithm implementation.} The implementation of the HHL algorithm can be divided in three subroutines: QPE to compute the eigenvalues of the problem matrix $A$, AQE to store the inverse of the eigenvalues in the amplitude of the ancilla qubit, and inverse QPE to reset the registers back to the ground state $|0\rangle^{\otimes n_R}_\text{R}$. {\bf (b) Multi-qubit decomposition of the AQE step for a ${\bf n_R=4}$ qubits example.} The rotations $V$ and $R_z\left(\phi\left(p\right)\right)$ performed on the ancilla qubit are defined in Eq. \eqref{eqV} and \eqref{eqRz}, respectively. {\bf (c) Binary reflected ${\bf 4}$-bit Gray code to define the positions of the control nodes.} The black and white rectangles refer to bit values one and zero, respectively. {\bf (d) AQE implementation using only single- and two-qubit gates.} Using the $4$-bit Gray code and following Ref. \citep{MVBS2004}, it is possible to decompose the multi-qubit AQE implementation into single-qubit rotations around the $z$-axis and CNOTs. The value of each $\theta_i$ follows from Eq. \eqref{eqThetaJ}. {\bf (e) Kite-like connectivity of the 	quantum processor.} The register qubits need to be connected among themselves, the ancilla qubit and the memory qubits, whereas the ancilla and the memory qubits need not to be connected directly.}
\label{Fig1}
\end{figure*}

The Harrow-Hassidim-Lloyd (HHL) quantum algorithm was proposed in 2009 \citep{HHL2009} to solve linear systems of equations: given a matrix $A$ and a vector $\vec{b}$, find a vector $\vec{x}$ such that $A\vec{x} = \vec{b}$. Due to the significance of linear systems of equations in several fields in science and engineering, this algorithm has attracted significant attention. However, one of the main issues with the HHL algorithm is, to our knowledge, that there is no explicit implementation for one of its key subroutines, known as ancilla quantum encoding (AQE). This subroutine stores the inversion of the eigenvalues of the problem matrix $A$, into the amplitude of an ancillary qubit.

In the literature, this routine has been circumvented by employing different tricks. For instance, by using a previous knowledge about the eigenvalues of the matrix $A$. Then, it is straightforward to tailor the best set of rotations to map those values into the amplitude of the ancilla qubit. Several experimental implementations in different experimental platforms \citep{Photonic_HHL_2013, NMH_HHL_2014, Photonic_HHL_2014, JWP2017} have followed this approach to solve $2\times 2$ linear system of equations. There is an alternative implementation of the HHL for which no spectral information of the matrix $A$ is required. In Ref. \citep{LJL2019} a quantum-classical hybridization of the algorithm is proposed. They repeatedly perform the quantum phase estimation (QPE) to obtain a $n_R$-bit description of the eigenvalues of $A$. Then, they determine the simplest circuit implementation to perform the AQE, tailored for those eigenvalues. Once the AQE is determined, it is possible to perform the complete HHL algorithm. However, this alternative fails if the vector $|b\rangle$ is not efficiently prepared. Despite the intrinsic interest of these approaches, both of them jeopardize the advantage of the algorithm, since they presume previous knowledge about the eigenvalues of the matrix $A$.

Algorithms such as the HHL are hard to implement in noisy intermediate-scale quantum (NISQ) devices. Indeed, current quantum computers are still not sufficiently robust against noise, which limits the depth of the algorithms that we can implement. These noise sources become much or less manageable depending on the computational paradigm. The most extended is the digital quantum computation (DQC) paradigm, in which the algorithm is decomposed into one- and two-qubit gates. Short depth quantum circuits have been implemented with this paradigm in several fields such as quantum machine learning \citep{ARSL2018, OSCSL2018}, finance \citep{YDING2019, PRICING2019, Javi2021}, open quantum systems \citep{SSSPS2016}, quantum chemistry \citep{GALMSSL2016}, or quantum field theories \citep{LSKDBLD2017}. In absence of quantum error correction, the main drawback of this paradigm is related to the cumulative noise that arises whenever a two-qubit gate is applied. 

To apply a two-qubit gate between qubits A and B, the natural interaction between them has to be enhanced,
while keeping the rest of interactions suppressed. If the interaction with the rest of the qubits of the system is not correctly attenuated, then the logical two-qubit operation is not rightly implemented, leading to errors in our quantum routine. Using quantum control techniques, it is possible to mitigate the error of the two-qubit gates, but those techniques are not scalable. Thus, it seems natural to get advantage of the intrinsic interaction in the processor as the resource to perform quantum computing, avoiding the interaction suppression which leads to errors. That is precisely the main idea behind the digital-analog quantum computation (DAQC) paradigm. The DAQC paradigm makes use of the natural interaction among the elements that conform a quantum system to perform quantum simulations, together with single-qubit gates to change the state of a particular qubit. Consequently, DAQC merges the flexibility of digital quantum computation with the robustness of analog simulations \citep{Parra2018, LPSS2018, MLSS2020, Headley2020, Celeri2021, GMS2021}.

A special case of two-qubit gate is the SWAP gate. This gate is not native in any quantum processor, thus it must be decomposed in terms of the intrinsic quantum gates, which is extremely expensive. Therefore, an adequate connectivity in the processor may dramatically reduce the use of SAWP gates, substantially decreasing the depth of the algorithm, which leads to an enhancement of the total fidelity. This is precisely the idea behind the co-design: build a quantum processor whose architecture is adapted to the connectivity of a concrete algorithm, simultaneously customizing the implementation of the algorithm in terms of the native interactions of the quantum platform.

In this Article, we propose a quantum processor architecture that is tailored for the HHL algorithm. More specifically, we introduce a systematic manner to implement the AQE subroutine which is independent of the problem matrix $A$ and, thus, it does not require the classical hybridization of the HHL algorithm. Due to its relevance, we focus on the computation of $f(\lambda) = 1/\lambda$, but this subroutine can perform any other function $f(A)$, as long as it satisfies the conditions specified in the supplementary material of Ref. \citep{HHL2009}. This implementation requires an exponential number of two-qubit gates which is very demanding for the NISQ era and hinders us from implementing the fully quantum algorithm in present technology. The found precision of the algorithm depends on the precision of the estimation of the eigenvalues given the number of register qubits $n_R$ and the precision of the estimation of the found mean values $\langle c | f(A) | b\rangle$, which depends on the number of samples $N_s$. These errors are independent, consequently, it is useless to reduce one of them far beyond the other, so $n_R$ can be bounded by $\mathcal{O}\left(\log\left[\kappa \sqrt{N_s}\right]\right)$ in terms of $N_s$. Consequently, the depth of the AQE step depends on the precision, but not on the dimension of the problem matrix $A$.

Taking into account the connectivity required by the algorithm, we propose a co-design ``kite-like" architecture for the quantum processor that reduces the number of SWAP gates, improving the implementation of the HHL algorithm in NISQ devices. Additionally, aiming at the experimental realization of the algorithm, we propose a digital-analog implementation in the kite-like quantum processor to perform the complete algorithm.

An outline for our work is as follows. In Sec. \ref{section_II} we present a theoretical description of the HHL algorithm divided into three steps, as well as the digital implementation of each step (Sec. \ref{section_IIa}). Additionally, we describe how the systematic implementation of the AQE step can be achieved for both the HHL and the concrete problem of solving linear system of equations (Sec. \ref{section_IIb}). Afterwards, in Sec. \ref{section_III}, we propose a co-designed quantum processor architecture that can enhance the performance of the HHL. Finally. in Sec. \ref{section_IV}, we present the DAQC description for the HHL algorithm. The conclusions are in Sec. \ref{section_V}.

\section{Fully quantum implementation of the HHL algorithm} \label{section_II}

In this section, we review the HHL algorithm and describe its implementation under the DQC paradigm. We go into details on how to apply the AQE step, so that the algorithm can be performed without previous knowledge of the eigenvalues of the problem matrix $A$.

For a $s$-sparse system matrix of size $N\times N$ and condition number $\kappa$, which is given by the ratio of the maximal and minimal singular values of $A$, the HHL algorithm can reach a desired computational accuracy $\epsilon$ within a running time of $\mathcal{O}\left(\log \left(N\right) s^2\kappa^2  \epsilon\right)$ under specific circumstances \citep{A2015}, comparing to the best known classical algorithm of $\mathcal{O}\left( N_s \kappa/\log \epsilon\right)$.

The algorithm involves three sets of qubits: a single ancilla qubit which stores the inverse of the eigenvalues of the matrix problem in its amplitude, a register of $n_R$ qubits to encode the $n_R$-binary representation of the eigenvalues of the problem matrix, and a set of $n_M= \log_2 (\dim |b\rangle)$ memory qubits used to load the state $|b\rangle$ and store the output $|x\rangle$. The amount of qubits required to perform the HHL algorithm depends both on the size of the matrix $A$ and the precision one would like to reach.

\subsection{Description of the HHL algorithm} \label{section_IIa}

The HHL algorithm is divided in three steps: QPE to compute the eigenvalues of the problem matrix $A$, AQE to map the inverse of those eigenvalues into the amplitude of the ancilla qubit, and inverse QPE to reset the registers back to the ground state $|0\rangle^{\otimes n_R}_R$. In the following, we describe each subroutine in detail. In Fig. \ref{Fig1}(a)  we show a scheme representation of the algorithm: the qubits are separated in the thee different sets and the steps are delimited in three colored boxes, one for each routine.

Since the HHL can be employed as a subroutine of a bigger problem, it is reasonable to assume that the memory qubits are initialize in the state $|b\rangle_\text{M} = \sum_{i=1}^N b_i|i\rangle$, where $|i\rangle$ denotes the computational basis of the $n_M$ qubits, as a consequence of previous operations in the system. If this was not the case and if $b_i$ and $\sum_i |b_i|^2$ are efficiently computable, then it is possible to prepare $|b\rangle$ following the procedure described in Ref. \citep{GR2002}. Either way, the system is initially in the state $|0\rangle_\text{A} \otimes |0\rangle_\text{R}^{\otimes n_R} \otimes |b\rangle_\text{M}$.

The first step of the algorithm is to apply QPE to compute the eigenvalues $\lambda_j$ of $A$ and encode them in a binary form into the state of the register qubits. Given $|b\rangle=\sum_{j=1}^N\alpha_j|u_j\rangle$, where $|u_j\rangle$ is the eigenvector basis of the matrix $A$, the state of the system after the QPE is $
|0\rangle_A\otimes
\sum_{j=1}^N\sum_{k=0}^{2^{n_R}-1} \alpha_j\beta_{k|j}|k\rangle_R
\otimes|u_j\rangle_M
$, where the coefficient $\beta_{k|j}|k$ is defined as
\begin{equation}
\beta_{k|j}=\frac{1}{2^{n_R}}\sum_{y=0}^{2^{n_R}-1}e^{2\pi i y \left( \lambda_j - k/2^{n_R} \right)}.
\end{equation}
If all the eigenvalues $\lambda_j$ are perfectly $n_R$-estimated, we can relabel them as $\widetilde{\lambda_k}\equiv k/2^{n_R}$. So that $\beta_{k|j}=\delta_{\lambda_k,\lambda_j}$, and the final state of the system after the QPE is performed is as follows,
\begin{equation}
|0\rangle_\text{A}\otimes
\sum_{j=1}^N\sum_{k=0}^{2^{n_R}-1} \alpha_j|\widetilde{\lambda_k}\rangle_\text{R}
\otimes|u_j\rangle_\text{M}
.
\end{equation} 

Once the estimated eigenvalues are encoded in a state superposition of the register qubits, the AQE maps them into the amplitude of the ancillary qubit, so that the resulting state of the ancilla is
\begin{equation}
\sum_{j=1}^N\sum_{k=0}^{2^{n_R}-1}\left(\sqrt{1-\frac{C^2}{\widetilde{\lambda_k^2}}}|0\rangle_\text{A}+\frac{C}{\widetilde{\lambda_k}}|1\rangle_\text{A}\right)\alpha_j|\widetilde{\lambda_k}\rangle|u_j\rangle,
\end{equation}
with $C \leqslant 1$ being a normalization constant chosen to be $\mathcal{O}\left(1/\kappa\right)$. In this work we propose a systematic and fully quantum protocol to achieve this mapping without needing to know in advance the eigenvalues of $A$. We describe the process within the next subsection. 

Finally, the inverse QPE has to be performed in order to uncompute the $|\lambda_j\rangle$ on the registers and reset them to the initial $|0\rangle^{\otimes n_R}$ state. After this step, the system is found in the state
\begin{equation}\label{EQ_final_state}
\sum_{j=1}^N\sum_{k=0}^{2^{n_R}-1}
\left(\sqrt{1-\frac{C^2}{\widetilde{\lambda_k^2}}}|0\rangle_\text{A}+\frac{C}{\widetilde{\lambda_k}}|1\rangle_\text{A}\right)\otimes|0\rangle_\text{R}\otimes \alpha_j |u_j\rangle_\text{M}
\end{equation}

To get the normalized solution of the linear equation, the ancillary qubit  has to be measure in the $Z$-axis. If the outcome state of the ancilla qubit is $|1\rangle_\text{A}$, then the state describing the system successfully represents the solution of the linear equation as
\begin{equation}\label{EQ_measurement}
\frac{1}{C}\sum_{j=1}^N\frac{\alpha_j}{\widetilde{\lambda_k}}|u_j\rangle_\text{M},
\end{equation}
up to a normalization factor. The pure state is obtained when the matrix $A$ is perfectly $n_R$-estimated. If there exists an eigenvalue of $A$ which is not perfectly $n_R$-estimated, then the total state of Eq. (\ref{EQ_final_state}) becomes a pure entangled state so that the state in Eq. (\ref{EQ_measurement}) turns into a mixed state.

\begin{figure}
\centering
\includegraphics[width=0.48\textwidth]{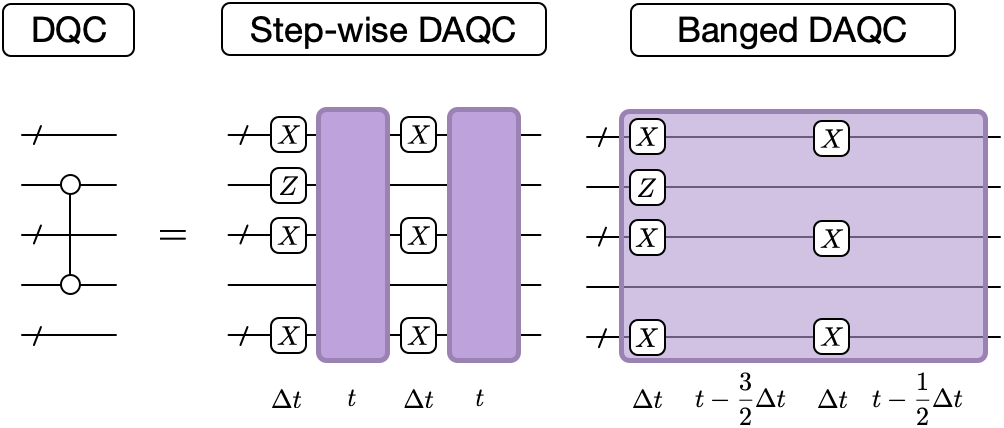}
\caption{{\bf DAQC decomposition of a $cZ$ gate.} To implement a $cZ$ gate using sDAQC techniques, two analog blocks of time $t=\pi/8$ are required. For the digital blocks, an $X$-gate must be applied before and after the first analog blocks on every qubit except for the qubits of the $cZ$-gate, where we will apply a $Z$-gate only in one of them, before or after the analog block. This freedom of choice comes from the fact the the interaction Hamiltonian commutes with the $Z$-gate. In the bDAQC case, the single-qubit rotations are applied on top of the interaction.
}
\label{Fig2}
\end{figure}

\begin{figure*}
\centering
\includegraphics[width=0.98\textwidth]{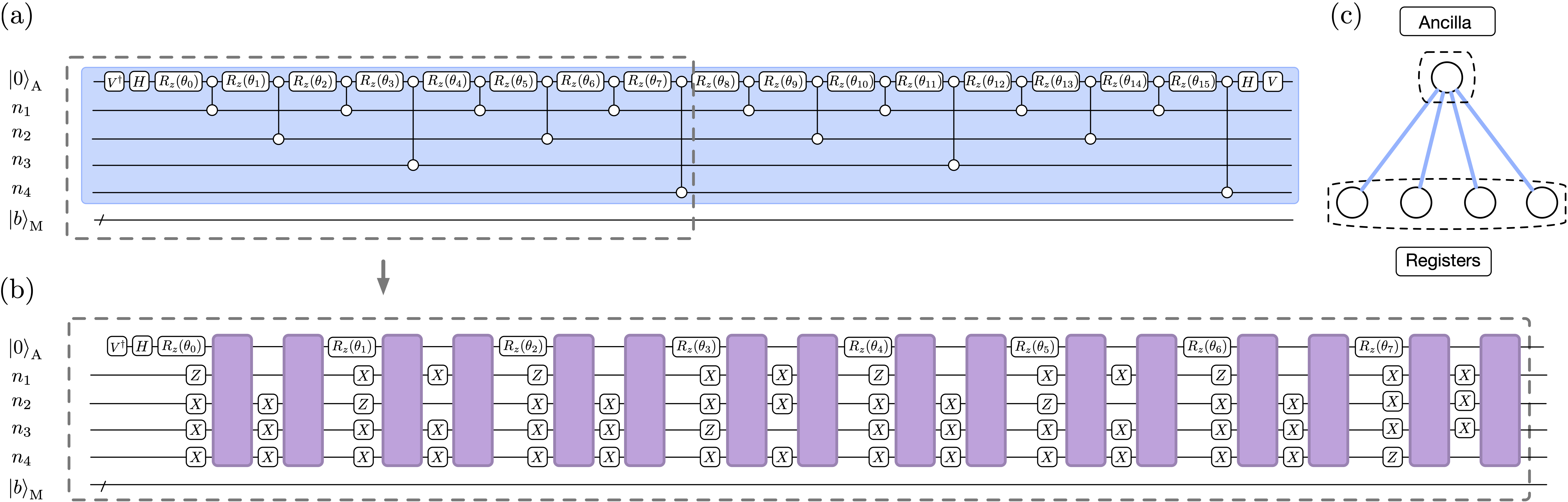}
\caption{{\bf Implementation of the AQE using DQC and DAQC techniques.} {\bf (a) DQC AQE decomposition into $R_z$ and $cZ$ gates} To build the DAQC protocol for the AQE, we first decompose this step into $cZ$ gates, rather that CNOT gates. {\bf (b) sDAQC AQE} Using as a resource the Ising Hamiltoninan that model the interaction between the qubits of the quantum processor, it is possible to decompose each $cZ$-gate on terms of single-qubit gates and analog blocks of time $t$. {\bf (c) Optimal connectivity of the ancilla and register qubits.}
}
\label{Fig3}
\end{figure*} 

\subsection{Ancilla quantum encoding} \label{section_IIb}

The AQE maps the superposition state of the registers into the amplitude of the ancilla qubit by means of the application of different controlled rotations. These rotation operations control each register qubit and act on the ancilla. In previous literature, the angle of these rotations was computed using the eigenvalues of the matrix $A$, which requires a previous knowledge of this information either before formulating the problem \citep{JWP2017}, or by applying QPE \citep{LJL2019}. But, if we know this information beforehand, then the problem of solving the linear system of equation becomes trivial.

In this section, we present a systematic way to perform the AQE operation without the necessity of knowing in advance further information about the problem matrix $A$. This way, our proposal constitute, to our knowledge, the first fully quantum explicit implementation of the HHL algorithm.

To perform the AQE step, we will first rotate the system applying the unitary gate $V$, which is defined as
\begin{equation}\label{eqV}
V = \frac{1}{\sqrt{2}}\begin{pmatrix}
-i & i\\
1 & 1
\end{pmatrix},
\end{equation}
and then, apply a series of rotations around the $Z$-axis in the ancilla qubit controlled by the state of the register qubits. This operation is defined as
\begin{equation}
U_{\text{AQE}} = \sum_{p=0}^{2^{n_R}-1} R_z\left(-\phi(p)\right)_\text{A} \otimes |\vec{b}_p\rangle\langle \vec{b}_p|_\text{R},
\end{equation}
where,
\begin{eqnarray}
R_z \left(\phi(p)\right) &=& \begin{pmatrix}
e^{-i\phi(p)/2} & 0\\
0 & e^{i\phi(p)/2}
\end{pmatrix}, \label{eqRz}\\
\phi(p) &=& \left\{ \begin{array}{lr}
0  & \text{if } p = 0\\
2 \arcsin\frac{1}{p} & \text{otherwise}
\end{array}\right. .
\end{eqnarray}
The bit-string $\vec{b}_p$ is the binary representation of the decimal number $p$, in other words, $p=\sum_{i=0}^{n_R} (\vec{b}_p)_i 2^{n_R-i}$. In Fig. \ref{Fig1}(b), we show an example of the AQE routine for a $n_R=4$ qubits case.

This implementation of the AQE step, requires an exponential number of multi-controlled gates. Following Ref. \citep{MVBS2004}, it can be decompose in $4^{n_R}$ single-qubit gates and $4^{n_R} - 2^{n_R+1}$ CNOT gates, which might be too demanding for NISQ devices. Therefore, the depth of the AQE step depends on the number of register qubits, and this number is related to the precision and error of the HHL algorithm.

In order to implement these multi-controlled operations employing only single- and two-qubit gates acting on the ancilla, we employ the procedure proposed in Ref. \citep{MVBS2004}. There, they provide an equivalent circuit which employs subsequent rotations with modified angles alternated with CNOT gates between the control and the target qubits. Consequently, the operation described in this section can be implemented with the circuit represented in Fig.\ref{Fig1}(d), where the angles $\theta_i$ are the modified angles of the single-qubit rotations related with the original angles by
\begin{equation}
\mqty(\theta_1 \\ \theta_2 \\ \vdots \\ \theta_{2^{n_R}})
=
M^{-1} 
\mqty(\phi (0) \\ \phi(1) \\ \vdots \\ \phi(2^{n_R}-1)),
\end{equation}
where
\begin{equation}
M_{ij} = (-1)^{\text{bin}(i-1) \cdot g(j-1)}
\text{ and }
M^{-1} = \frac{1}{2^{n_R}} M^{T},
\end{equation}
where $\text{bin}(i)$ is the $n_R$-bit binary representation of the integer $i$, $g(j)$ represents the $j$-th string of the binary reflected Gray code (counting from 0), and the dot represents the bit-wise product. Hence, substituting the expression for $\phi(b)$ we obtain
\begin{equation}\label{eqThetaJ}
\theta_i = \frac{1}{2^{n_R}} \sum_{j=1}^{2^{n_R}-1} (-1)^{\text{bin}(j) \cdot g(i-1)} \arcsin \frac{1}{j}.
\end{equation}

The expected error of the algorithm in terms of the final state is $|| |x\rangle - |\tilde{x}\rangle || = \mathcal{O}\left(\kappa/2^{n_R}\right)$, where $|x\rangle$ is the solution of the $A|x\rangle=|b\rangle$ problem and $|\bar{x}\rangle$ is the solution obtained by the algorithm. We now consider the scenario in which we are interested in estimating the expected value of an observable $\Omega$ with a limited number of samples $N_s$. The error scales as $\mathcal{O}\left( 1/\sqrt{N_s}\right)$. Thus, the expected error, $\epsilon$, of estimating an observable $\Omega$ in $|x\rangle$ by measuring $|\tilde{x}\rangle$ is
\begin{equation}\label{eq12}
\epsilon = \mathcal{O}\left(\frac{1}{\sqrt{N_s}} +  \frac{\eta}{2^{n_R}}\right),
\end{equation}
where $\eta$ is a constant of $\mathcal{O}(\kappa)$, being $\kappa$ the condition number of the problem matrix $A$. If the number of  samples is fixed in $N_s$ samples, then, the number of register qubits can be estimated by imposing that none of the summands in Eq. \eqref{eq12} is dominant. This implies that the sensible amount of register qubits $n_R$ is of $\mathcal{O}\left(\log\left[\kappa \sqrt{N_s}\right]\right)$, which only depends on the number of samples, but not on the size of the problem matrix $A$.

In the {\it generalized}-HHL (gHHL), one wants to calculate $\vec{x} = f(A) \vec{b}$, where $f(A)$ is a matrix function. Then, our method can easily be adapted to fulfill this task. It is straightforward to show that the desired operation requires the AQE to perform the following operation on the ancilla qubit,
\begin{equation}
|0\rangle_\text{A} |\widetilde{\lambda_k}\rangle_\text{R} \xrightarrow{\text{AQE}} \left(\sqrt{1-f(\lambda^2)}|0\rangle_\text{A} + f(\lambda)|1\rangle_\text{A}\right)|\widetilde{\lambda_k}\rangle_\text{R}.
\end{equation}
To achieve this mapping, we would redefine the rotation angles of the multi-controlled operations as $\phi(p) = 2 \arcsin f(p/2^{n_R})$. However, the found bounds for the error are only satisfied if the function $f(\lambda)$ satisfies the condition
\begin{equation}
\left| \left( \frac{\text{d}f}{\text{d}x}\right)^2 \left(1+\frac{1}{1-f(x)^2}\right)\right|<\eta^2 \quad \text{in the interval } x\in [0,1),
\end{equation}
that can be derived form the Lemma 2 of the supplementary material of Ref. \citep{HHL2009}, where $\eta$ is an arbitrary constant of order $\mathcal{O}(1)$. This condition on the function $f(\lambda)$ implies that, in a realistic case, one would construct an auxiliary function $F$ that satisfy that bound and behaves as $f(\lambda)$ in a certain interval, which is a similar procedure to normalizing a vector to load it as a quantum state.

\section{Co-designed processor architecture for the HHL} \label{section_III}

If the architecture of the processor in which the algorithm is being implemented is not optimal, a significant amount of SWAP gates might be required in order to replaced the missed connections. This demand increases the depth of the quantum circuit, making its implementation in NISQ devices a challenging task. Having an optimized quantum processor architecture keeps the depth controlled and reduces the amount of SWAP gates required to replace the missed connections. Here, we present an architecture for the quantum processing unit to implement the gHHL algorithm.

An optimized quantum processor architecture for the gHHL takes into account the dependence between the three sets of qubits in the different steps of the algorithm. Each set plays a different role in the algorithm and thus, require different connections: the register qubits are connected among them and, simultaneously, to the ancilla and memory qubits, whereas the ancilla and the memory sets need not to be connected directly. In Fig.\ref{Fig1} (e) we show how this leads to a ``kite-like'' architecture that satisfy the demanded connections of the qubit sets in the different steps of the gHHL.

In the context of hardware implementation, it is worth noting that we are speaking about logical qubits. This means that, each logical qubit can be constituted by a group of physical qubits on which we perform the necessary computational operations.

\section{Digital-analog implementation for the HHL algorithm}  \label{section_IV}

The DAQC paradigm combines analog blocks with digital steps to approximate any unitary with arbitrary precision. The digital steps are single-qubit gates and the analog blocks are constituted by the time evolution of the interaction Hamiltonian inherent to the quantum processor. By getting advantage of the natural interaction among qubits, the DAQC claims to be more resilient against noise than the fully digital one \citep{MLSS2020, GMS2021}. Here, we propose a DAQC implementation that takes into account all of the subroutines constituting the gHHL.

As previously mentioned, it is possible to describe the gHHL in a succession of three steps (QPE, AQE and inverse QPE). At the same time, QPE can be described as well by two subroutines: the controlled-Hamiltonian evolution and the inverse quantum Fourier transform. The division of each step into more fundamental subroutines simplifies the description of the complete gHHL algorithm in the DAQC paradigm, since some of those subroutines has already been worked out in previous works \citep{LPSS2018, MLSS2020}.

To perform the controlled-Hamiltonian evolution that constitutes the first part of the QPE, we assume that the problem matrix $A$ admits an $M$-body decomposition, where $M<<\log_2 \dim (A)$. Under this condition, it is possible to upload that matrix efficiently using the DAQC protocol as described in Ref. \citep{Parra2018}. The next step that conforms the QPE subroutine is the implementation of the inverse quantum Fourier transform on the register  qubits. This step was explicitly described in Ref. \citep{MLSS2020}. 

Finally, in order to implement the AQE step in a digital-analog paradigm, it is useful to firstly decompose it into single-qubit rotations, namely Hadamard gates and rotations around the X-axis, and controlled-Z ($cZ$) gates, as shown in Fig. \ref{Fig3} (a). Then, we can get a step-wise DAQC (sDAQC) decomposition by constructing the Hamiltonian of the $cZ$ gates, which is a two-body Ising Hamiltonian of the form
\begin{equation}
H_\text{cZ} = -\frac{1}{2}Z\otimes\mathbb{1}+\frac{1}{4} Z\otimes Z.
\end{equation}
In the sDAQC protocol, the interaction among the qubits of the system are switched off before applying the digital steps, and switched on again immediately afterwards. This approach does not substantially improve the result of the digital approach since the errors induced by the attenuation of the natural interaction between the qubits are similar. A better alternative is the banged DAQC (bDAQC). In this paradigm, the interaction between the qubits is not turned off to perform the single-qubit gates, which are applied on top of the interaction Hamiltonian. Although this means that an intrinsic error is introduced, this error scales better with the number of qubits when the time for the single-qubit gates, $\Delta t$, is significantly smaller that the natural time scale of the analog blocks. In Fig. \ref{Fig2}, we show an scheme of both the sDAQC and bDAQC protocols for the $cZ$ gate.

For simplicity, we will assume that the interaction between the qubits of our kite-like quantum processor is an homogeneous two-body Ising Hamiltonian, $H_{I}$. Then, every $cZ$ performed between the ancilla and one of the register qubits can be decomposed into two analog blocks in the following way
\begin{eqnarray}
U_\text{cZ} &=& e^{i\frac{\pi}{4} Z_\text{A}\otimes Z_\text{R}} \nonumber \\
&=& \left( \bigotimes_{\substack{k=1\\ k\neq i}}^{n_R}X_k\right) e^{i\frac{\pi}{8}H_I}  \left( \bigotimes_{\substack{k=1\\ k\neq i}}^{n_R}X_k\right) e^{i\frac{\pi}{8}H_I}.
\end{eqnarray}
In Fig. \ref{Fig2}, we show a scheme of the sDAQC implementation of a $cZ$ gate.

To implement the complete AQE using DAQC, we would decompose each $cZ$ in the digital and analog blocks described above. Each analog blocks would take the same amount of time $t$ to act, and the digital steps would take a fixed time $\Delta t$ to be applied. In Fig. \ref{Fig3} (b), we show the sDAQC implementation of the AQE. Only half of the procedure is represented since the other half is symmetrical. As we mention previously, in terms of scalability it is preferable to follow the bDAQC approach. For the sake of completeness, the explicit DAQC description of the AQE step comprises, at most, $2^{n_R}$ analog blocks and $(n_R+1) 2^{n_R}+1$ single-qubit gates. This description can be improved by employing optimization techniques.

\section{Conclusions} \label{section_V}

The implementation of the gHHL algorithm can be divided into three steps: QPE to compute the eigenvalues of the problem matrix $A$, AQE to map a function of the problem matrix $A$ into the amplitude of the ancillary qubit, and the inverse QPE to set the registers back to the ground state. The manner to perform the first and last steps is well described in the literature \citep{NCh2000}, whereas an explicit description of the AQE step has remained as an open question. In this work, we have shown how the AQE can be accomplished in a systematic manner which does not require the hybridization of the algorithm, making this the first fully quantum implementation of the gHHL, to our knowledge.

Implementing the AQE routine using our protocol requires an exponential number of single- and two-qubit gates, what has hindered us from implementing it in NISQ devices. This number depends on the amount of register qubits of our setup and is related to the precision of the algorithm. Attending to the number of samples required for a particular experiment, it is possible to diminish the number of register qubits needed to perform the gHHL algorithm, more concrete, for a $N_s$-sample experiment, the number of register qubits needed would be of order $\mathcal{O}\left(\log\left[\kappa\sqrt{N_s}\right]\right)$. This means that the number of register qubits $n_R$ depends exclusively on the bond for $f(A)$ and the number of samples $N_s$, but it is independent form the dimension of the matrix $A$.

In order to reduce the depth of the gHHL algorithm, we proposed a co-designed quantum processor tailored to the implementation of the gHHL. This structure significantly reduces the amount of SWAP gates needed to artificially connect the qubits of the system which are not physically connected but they are necessary to perform the algorithm. Thus, the co-design also reduces the depth of the algorithm. The kite-like quantum processor connects the register qubits with both the ancilla and the memory qubits, while keeping these last two sets of qubits disconnected from each other. 

There is another approach to the implementation of the gHHL that naturally arises from the idea of getting advantage of the connections naturally present in the quantum processor, and it is to implement the algorithm using DAQC techniques. In view of the results obtained in previous works about the simulation of quantum subroutines, such as the quantum Fourier transform, which forms part of the gHHL, it is expected that a DAQC implementation of the gHHL will produce better results in terms of noise scaling with the number of qubits \citep{MLSS2020, GMS2021}. For this reason, we propose a complete protocol for the implementation of the gHHL algorithm, which allows obtaining results when applied in a system with a high number of qubits.

\section{Acknowledgments}
The authors acknowledge financial support from the QUANTEK project from ELKARTEK program (KK-2021/00070), Spanish Ram\'on y Cajal Grant RYC-2020-030503-I and the project grant PID2021-125823NA-I00 funded by MCIN/AEI/10.13039/501100011033 and by “ERDF A way of making Europe” and "ERDF Invest in your Future", as well as from QMiCS (820505) and OpenSuperQ (820363) of the EU Flagship on Quantum Technologies, and the EU FET-Open projects Quromorphic (828826) and EPIQUS (899368). IQM Quantum Computers funded project ``Generating quantum algorithms and quantum processor optimization". R.I. acknowledges the support of the Basque Government Ph.D. grant PRE\_2021-1-0102.


\begin{thebibliography}{X}
\bibitem{HHL2009}
A. W. Harrow, A. Hassidim, and S. Lloyd, {\it Quantum Algorithm for Linear Systems of Equations}. Phys. Rev. Lett. {\bf 103}, 150502 (2009).

\bibitem{Photonic_HHL_2013}
X. D. Cai, C. Weedbrook, Z. E. Su, M. C. Chen, M. Gu, M. J. Zhu, L. Li, N. L. Liu, C. Y. Lu, and J. W. Pan, {\it Experimental quantum computing to solve systems of linear equations}. Phys. Rev. Lett. {\bf 110}, 230501 (2013).

\bibitem{NMH_HHL_2014}
J. Pan, Y. Cao, X. Yao, Z. Li, Ch. Ju, H. Chen, X. Peng, S. Kais, and J. Du, {\it Experimental realization of quantum algorithm for solving linear system of equations}. Phys. Rev. A {\bf 889}, 022313 (2014).

\bibitem{Photonic_HHL_2014}
S. Barz, I. Kassal, M. Ringbauer, Y. O. Lipp, B. Dakić, A. Aspuru-Guzik, and P. Walther, {\it A two-qubit photonic quantum processor and its application to solving systems of linear equations}. Sci. Rep. {\bf 4}, 6115 (2014).

\bibitem{JWP2017}
Y. Zheng, C. Song, M. Chen, B. Xia, W. Liu, Q. Guo, L. Zhang, D. Xu, H. Deng, K. Huang, Y. Wu, Z. Yan, D. Zheng, L. Lu, J. Pan, H. Wang, C. Lu, and X. Zhu, {\it Solving Systems of Linear Equations with a Superconducting Quantum Processor}. Phys. Rev. Lett. {\bf 118}, 210504 (2017).

\bibitem{LJL2019}
Y- Lee, J. Joo, and S. Lee, {\it Hybrid quantum linear equation algorithm and its experimental test on IBM Quantum Experience}. Sci. Rep. {\bf 9} 4778 (2019).

\bibitem{ARSL2018}
F. Albarr\' an-Arriagada, J. C. Retamal, E. Solano, and L. Lamata, {\it Measurement-based adaptation protocol with quantum reinforcement learning}. Physical Review A {\bf 98}, 042315 (2018).

\bibitem{OSCSL2018}
J. Olivares-S\' anchez, J. Casanova, E. Solano, and L.~Lamata, {\it Measurement-based adaptation protocol with quantum reinforcement learning in a Rigetti quantum computer}. arxiv:1811.07594 (2018).

\bibitem{YDING2019}
Y. Ding, L. Lamata, M. Sanz, J. D. Mart\' in-Guerrero, E. Lizaso, S. Mugel, X. Chen, R. Or\' us and E. Solano, {\it Towards Prediction of Financial Crashes with a D-Wave Quantum Computer}. arxiv:1904.05808 (2019).

\bibitem{PRICING2019}
A. Martin, B. Candelas, \'A. Rodr\'iguez-Rozas, J. D. Martín-Guerrero, X. Chen, L. Lamata, R. Orús, E. Solano, and M. Sanz, {\it Towards Pricing Financial Derivatives with an IBM Quantum Computer}. Pyhs. Rev. Research {\bf 3}, 013167 (2021).

\bibitem{Javi2021}
J. Gonzalez-Conde, \'A. Rodr\'iguez-Rozas, E. Solano, and M. Sanz, {\it Simulating option price dynamics with exponential quantum speedup}. arxiv: 2101.04023 (2021).


\bibitem{SSSPS2016}
R. Sweke, M. Sanz, I. Sinayskiy, F. Petruccione, and E. Solano, {\it Digital quantum simulation of many-body non-Markovian dynamics}. Physical Review A {\bf 94}, 022317 (2016).

\bibitem{GALMSSL2016}
L. Garc\' ia-\' Alvarez, U. Las Heras, A. Mezzacapo, M. Sanz, E.~Solano, and L. Lamata, {\it Quantum chemistry and charge transport in biomolecules with superconducting circuits}. Scientific Reports {\bf 6}, 27836 (2016).

\bibitem{LSKDBLD2017}
N. K. Langford, R. Sagastizabal, M. Kounalakis, C. Dickel, A.~Bruno, F. Luthi, D. J. Thoen, A. Endo, and L. DiCarlo, {\it Experimentally simulating the dynamics of quantum light and matter at deep-strong coupling}. Nature Communications {\bf 8}, 1715 (2017).

\bibitem{Parra2018}
A. Parra-Rodriguez, P. Lougovski, L. Lamata, E. Solano, and M. Sanz, {\it Digital-Analog Quantum Computation}. Physical Review A {\bf 101}, 022305 (2020).

\bibitem{LPSS2018} 
L. Lamata, A. Parra-Rodriguez, M. Sanz, and E. Solano, {\it Digital-analog quantum simulations with superconducting circuits}. Advances in Physics: X {\bf 3} 1457981 (2018).

\bibitem{MLSS2020} A. Martin, L. Lamata, E. Solano, and M. Sanz, \textit{Digital-analog quantum algorithm for the quantum Fourier transform}. Physical Review Research \textbf{2}, 013012 (2020).

\bibitem{Headley2020} 
D. Headley, T. Müller, A. Martin, E. Solano, M. Sanz, and F. K. Wilhelm, {\it Approximating the Quantum Approximate Optimisation Algorithm}. arxiv:2002.12215 (2021).

\bibitem{Celeri2021}
L. C. Céleri, D. Huerga, F. Albarrán-Arriagada, E. Solano, and M. Sanz, {\it Digital-analog quantum simulation of fermionic models}. arxiv: 2103.15689 (2021).

\bibitem{GMS2021}
P. Garc\'ia-Molina, A. Martin, and M. Sanz, {\it Noise in Digital and Digital-Analog Quantum Computation}. arxiv: 2107.12969.

\bibitem{A2015}
AAronson, S. {\it Read the fine print}. Nature Phys {\bf 11}, 291–293 (2015).


\bibitem{GR2002}
L. Grover, and T. Rudolph, {\it Creating superpositions that correspond to efficiently integrable probability distributions}. arxiv: 0208112 (2002).

\bibitem{MVBS2004} 
M. M\"{o}tt\"{o}nen, J. J. Vartiainen, V. Bergholm, and M. M. Salomaa, {\it Quantum circuits for general multi-qubit gates}. Phys. Rev. Lett. {\bf 96}, 130502 (2004).

\bibitem{NCh2000}
M. A. Nielsen, and I. L. Chuang, {\it Quantum information and quantum computation}. Cambridge: Cambridge University Press, 2(8), 23.

\end{thebibliography}
\end{document}